\documentclass[prb,fleqn,12pt,showpacs]{revtex4}
\usepackage{epsfig}
\usepackage{graphicx}
\usepackage{amsfonts}
\begin{document} 
\title{An effective quantum parameter \\ for strongly correlated metallic ferromagnets}
\author{Bhaskar Kamble}
\email{kbhaskar.iitk@gmail.com}
\affiliation{16 Aangan Apartments, Vastrapur, Ahmedabad - 380015}
\author{Avinash Singh}
\email{avinas@iitk.ac.in}
\affiliation{Indian Institute of Technology Kanpur - 208016}
\begin{abstract}
The correlated motion of electrons in multi-orbital metallic ferromagnets is investigated in terms of a realistic Hubbard model with ${\cal N}$-fold orbital degeneracy and arbitrary intra- and inter-orbital Coulomb interactions $U$ and $J$ using a Goldstone-mode-preserving non-perturbative scheme. An effective quantum parameter $'\hbar'=\frac{U^2+({\cal N}-1)J^2}{(U+({\cal N}-1)J)^2}$ is obtained which determines, in analogy with $1/S$ for quantum spin systems and $1/N$ for the $N$-orbital Hubbard model, the strength of correlation-induced quantum corrections to magnetic excitations. The rapid suppression of this quantum parameter with Hund's coupling $J$, especially for large ${\cal N}$, provides fundamental insight into the phenomenon of strong stabilization of metallic ferromagnetism by orbital degeneracy and Hund's coupling. This approach is illustrated for the case of ferromagnetic iron and the half metallic Heusler alloy $\rm Co_2 Mn Si$. For realistic values for iron, the calculated spin stiffness and Curie temperature values obtained are in quantitative agreement with measurements. Significantly, the contribution of long wavelength modes is shown to yield a nearly $\sim25\%$ reduction in the calculated Curie temperature. Finally, an outline is presented for extending the approach to generic multi-band metallic ferromagnets including realistic band-structure features of non-degenerate orbitals and inter-orbital hopping as obtained from LDA calculations. 
\end{abstract}
\maketitle
\newpage
\section{Introduction}

Dramatic improvements in experimental techniques such as Angle Resolved Photoemission Spectroscopy (ARPES)\cite{arpes} have led to important insight into the origin and role of correlation effects in itinerant ferromagnets such as iron, highlighting the coupling of electrons with magnons as playing a major role in the electron self energy renormalization and scattering rates. Correlation effects also play an important role in the observed zone boundary magnon softening and damping in ultrathin films of iron, as observed in recent Spin Resolved Electron Energy Loss Spectroscopy (SPEELS) experiments,\cite{speels} where the zone boundary magnon energies are much lower than those predicted within the random phase approximation (RPA), and the magnon energies were observed to depend non-monotonically on the film thickness.\cite{speels2} Further evidence of correlation effects is provided by {\it ab-initio} band structure calculations of half-metallic Heusler alloys showing emergence of non-quasiparticle (NQP) minority-spin states near the Fermi energy at finite temperatures, which has been suggested to be responsible for the strong suppression of tunneling magnetoresistance (TMR) ratio with temperature in ${\rm Co_2MnSi}$-based magnetic tunneling junctions (MTJ) observed in tunneling conductance measurements.\cite{chioncel_PRL_2008} These experiments conclusively highlight the importance of incorporating electron-magnon coupling effects in the correlated electron spin dynamics in metallic ferromagnets.

A considerable amount of work has been devoted to understanding the electronic band structure of these metallic ferromagnets,\cite{barriga_2009} starting from the Local Spin Density Approximation (LSDA) within the Density Functional Theory (DFT),\cite{LSDA_DFT} which accounted for correlations only in a limited way. The development of several extensions such as the LDA+U, LDA++, and LDA+DMFT,\cite{anisimov_1991,LDA++,lichtenstein_PRL_2001} has led to considerable progress in incorporating correlation effects in realistic band structure calculations. However, here the correlation term is incorporated either at the mean-field level or within a local self-energy approximation which neglects the momentum dependence. These methods therefore cannot be used to directly address spin wave excitations as they do not explicitly preserve the spin rotation symmetry, for which vertex corrections must also be included systematically,\cite{hertz-edwards1} and also predict much higher Curie temperatures\cite{lichtenstein_PRL_2001} than observed experimentally due to neglect of long wavelength spin fluctuation modes. 

Since metallic ferromagnets are characterized by intermediate to strong correlations, a proper description of spin waves must incorporate correlation effects non-perturbatively and simultaneously preserve the Goldstone mode. There are mainly two theoretical approaches for studying spin-wave excitations in itinerant ferromagnets --- the random phase approximation (RPA),\cite{rpa_bulk,rpa_thinfilms,naito_JPSJ_2007} and mapping to an equivalent Heisenberg model of localized spins by using the magnetic force theorem and its generalization to compute the exchange interaction parameters.\cite{localized1,localized3,thoene_2009} Due to neglect of correlation effects, the RPA is well known to overestimate the spin stiffness, magnon energies, and stability of the ferromagnetic state.\cite{hertz-edwards1} On the other hand, mapping to an  effective Heisenberg model does not capture typically itinerant features such as zero-temperature magnon damping. The adiabatic approximation has been used to investigate spin dynamics of ultra-thin films,\cite{localized1} but it has been pointed out that this approach breaks down for large wave-vector modes.\cite{rpa_thinfilms} The localized spin model has also proved unsatisfactory in explaining the doping dependence of the anomalous softening and damping of zone boundary spin wave modes in the CMR manganites.\cite{zhang_JPCM_2007} Although Linear Response Density Functional Theory (LRDFT)-based studies of spin dynamics of iron\cite{buczek_JMMM_2010,savrasov_1998} and the Heusler alloys\cite{buczek} account for damping of high-energy magnon modes due to decay into Stoner excitations, it has been pointed out that spin-charge coupling in a band ferromagnet results in significant magnon damping for modes lying even within the Stoner gap.\cite{sp_2008} 

In this situation, it will be useful to have a scheme which could incorporate features of the realistic electronic band structure and simultaneously take into account the most important correlation effects within a non-perturbative and Goldstone mode preserving approach. Such a scheme would also be useful from the technological point of view, since many half-metallic ferromagnets such as the Heusler alloys are being intensively investigated due to their potential applications in the spintronics industry.

Recently, correlation effects in metallic ferromagnets have been investigated using a non-perturbative, inverse-degeneracy based expansion scheme in which self energy and vertex corrections are included systematically so that the spin rotation symmetry and the Goldstone mode are explicitly preserved order by order.\cite{AS_2006,pandey_PRB_2007,sp_2008}
Within this approach, the enhancement of ferromagnetism due to suppression of correlation-induced quantum corrections was found to be strongly dependent on several electronic band features such as lattice-type, dimensionality, $t'$-induced DOS asymmetry, and band filling. The magnon self-energy was also investigated in the context of zone boundary magnon softening and magnon damping observed in manganites and ultra-thin transition-metal films.\cite{sp_2008,dksin_PRB_2010,qfklm2} 

The ${\cal N}$-orbital Hubbard model considered in the above works involved the orbitally-symmetric case with identical intra-orbital and inter-orbital Coulomb interactions. In this paper, we shall extend this spin-rotationally symmetric Goldstone-mode preserving approach to the general orbitally-asymmetric case with arbitrary intra- and inter-orbital Coulomb interactions. This provides an important extension of our recent work on quantum corrections in the doubly-degenerate (${\cal N}=2$) Hubbard model with intra-orbital Coulomb interaction $U$ and inter-orbital Hund's exchange $J$.\cite{bhaskar} We will derive an effective quantum parameter $'\hbar'=(U^2+({\cal N}-1)J^2)/(U+({\cal N}-1)J)^2$ which determines, in analogy with $1/{\cal N}$ for the orbitally-symmetric Hubbard model and $1/S$ for quantum spin systems, the strength of the correlation-induced quantum corrections in a realistic multi-orbital metallic ferromagnet.

As an illustration of this spin-rotationally invariant approach for investigating correlation effects in multi-band metallic ferromagnets, we will calculate electronic and magnetic properties in a five-orbital model with realistic parameters corresponding to ferromagnetic iron. The inclusion of long wavelength spin-fluctuation modes in the finite-temperature spin dynamics is distinct advantage of our approach, and it provides a quantitative measure of the Curie-temperature overestimate in local self-energy based calculations (for example, the LDA+DMFT) which neglect contribution of long wavelength modes. We will also outline an extension to a realistic multi-band model including non-degenerate orbitals and inter-orbital hopping as typically obtained in a LDA calculation.  

The outline of this paper is as follows. After introducing the orbitally-asymmetric ${\cal N}$-orbital Hubbard model in Section II, the transverse spin fluctuation propagator is studied in section III, and the effective quantum parameter is derived from the first order quantum correction diagrams for the irreducible particle-hole propagator obtained here. Our approach is illustrated with two applications corresponding to ferromagnetic iron in Section IV and the Heusler alloy ${\rm Co_2MnSi}$ in Section V. The extension to realistic multi-band model including non-degenerate orbitals and inter-orbital hopping is outlined in Section VI, and our conclusions are presented in section VII.

\section{${\cal N}$-orbital Hubbard model with Hund's coupling}
We consider the following orbitally asymmetric ${\cal N}$-orbital Hubbard model with arbitrary intra-orbital ($U$) and inter-orbital $(J)$ Coulomb interactions:
\begin{equation}
{\cal H}=-\sum_{\langle ij \rangle,\sigma,\mu}t_{ij} a^{\dagger}_{i\sigma\mu} a_{j\sigma\mu} - U\sum_{i,\mu}{\bf S}_{i\mu}\cdot{\bf S}_{i\mu}-J\sum_{i,\mu\neq\nu}{\bf S}_{i\mu}\cdot{\bf S}_{i\nu},
\end{equation}
where $\mu$ and $\nu$ refer to the ${\cal N}$ degenerate orbitals at each lattice site $i$, and ${\bf S}_{i\mu}=\psi^{\dagger}_{i\mu}({\mbox{\boldmath $\sigma$}}/2)\psi_{i\mu}$ is the local electron spin operator for the $\mu$ orbital in terms of the corresponding fermion operators $\psi^{\dagger}_{i\mu}=(a^{\dagger}_{i\uparrow\mu}\hspace{3mm}a^{\dagger}_{i\downarrow\mu})$ and the Pauli matrices ${\mbox{\boldmath $\sigma$}}$. For $J=U$, the model reduces to the orbitally symmetric case considered earlier.\cite{AS_2006} The inter-orbital density interaction term $Vn_{i\mu}n_{i\nu}$ is not included here as this charge term has no leading order effect on magnetism and only weak effects on spin dynamics when quantum corrections are included away from the onset of staggered orbital ordering. The role of orbital fluctuations on spin dynamics due to this inter-orbital density interaction term has been studied recently in the context of the observed zone-boundary anomalies in manganites.\cite{dksin_PRB_2010,qfklm2}

The continuous spin rotation symmetry of the above Hamiltonian implies the existence of Goldstone modes in the spontaneously broken symmetry state. In the following we will present a non-perturbative scheme in which the correlation-induced self energy and vertex corrections are incorporated systematically so that the Goldstone mode is explicitly preserved order by order.

\section{transverse spin fluctuations}
We assume a ferromagnetic ground state with magnetization in the $z$ direction and examine transverse spin fluctuations representing both collective (spin-wave) and single-particle (Stoner) excitations. We consider the time-ordered transverse spin-fluctuation propagator in this broken-symmetry state:
\begin{equation}
\chi^{-+}_{\mu\nu}({\bf q},\omega)=i\int dt e^{i\omega(t-t')}\sum_{j}e^{i{\bf q}.({\bf r}_i-{\bf r}_j)}\langle\Psi_0|T[S^{-}_{i\mu}(t)S^{+}_{j\nu}(t')]|\Psi_0\rangle,
\label{one}
\end{equation}
where $\mu$ and $\nu$ are any of the ${\cal N}$ orbital indices, and the fermion-spin lowering and spin-raising operators $S^{\mp}_{i\mu}=\psi^{\dagger}_{i\mu}(\sigma^{\mp}/2)\psi_{i\mu}$. 

In terms of the irreducible particle-hole propagator $\phi_{\mu\nu}({\bf q},\omega)$, the spin-fluctuation propagator can be written exactly as:
\begin{equation}
\chi_{\mu\nu}^{-+}({\bf q},\omega)=\phi_{\mu\nu}({\bf q},\omega)+\phi_{\mu\mu'}({\bf q},\omega)U_{\mu'\nu'}\chi_{\mu'\nu'}^{-+}({\bf q},\omega),
\label{c3e15}
\end{equation}
where the interaction term $U_{\mu\nu}=U$ for $\mu=\nu$, and $U_{\mu\nu}=J$ for $\mu \neq \nu$, and summation over repeated indices is implied. It is physically relevant to consider the total transverse spin-fluctuation propagator:
\begin{equation}
\chi^{-+}({\bf q},\omega) = \sum_{\mu}\chi^{-+}_{\alpha\mu}({\bf q},\omega),
\label{c3e16}
\end{equation}
as it measures the response to external probes such as magnetic field or neutron magnetic moment which couple equally to the electron moment for all orbitals. It is particularly convenient to solve the coupled equations then to obtain:
\begin{equation}
\chi^{-+}({\bf q},\omega) = \frac{\phi ({\bf q},\omega)}{1-U^+ \phi ({\bf q},\omega)},
\label{c3e18}
\end{equation}
where the interaction term $U^+ = U+({\cal N}-1)J$, and the total irreducible particle-hole propagator:
\begin{equation}
\phi({\bf q},\omega) = \sum_{\mu}\phi_{\alpha\mu}({\bf q},\omega).
\label{c3e17}
\end{equation}

In analogy with the $1/{\cal N}$ expansion for the orbitally symmetric ${\cal N}$-orbital Hubbard model, we consider a systematic expansion:
\begin{equation}
\phi=\phi^{(0)}+\phi^{(1)}+\phi^{(2)}+\cdots
\end{equation}
for the irreducible propagator $\phi({\bf q},\omega)$ in terms of fluctuations. The first term $\phi^{(0)}$ is simply the bare particle-hole propagator, whereas the higher-order terms $\phi^{(1)}$, $\phi^{(2)}$ etc. represent correlation-induced quantum corrections involving self energy and vertex corrections. 

\subsection{Random Phase Approximation}
Retaining only the zeroth-order term $\phi^{(0)}$ in the expansion yields the random phase approximation, amounting to a ``classical-level'' description of non-interacting spin-fluctuation modes. As the hopping term is diagonal in orbital indices, the zeroth-order term involves only the intra-orbital contribution:
\begin{equation}
\phi^{(0)}_{\alpha\alpha}({\bf q},\omega) \equiv \chi_0({\bf q},\omega)=\sum_{\bf k}\frac{1}{\epsilon^{\downarrow+}_{{\bf k}-{\bf q}}-\epsilon^{\uparrow-}_{\bf k}+\omega-i\eta},
\end{equation}
where the Hartree-Fock level band energies $\epsilon^\sigma_{\bf k}=\epsilon_{\bf k}-\sigma\Delta$ involve the exchange splitting:
\begin{equation}
2\Delta=[U+({\cal N}-1)J]m
\end{equation}
between the two spin bands. The superscripts $+$ ($-$) refer to particle (hole) states above (below) the Fermi energy $\epsilon_F$. Here the magnetization $m=2\langle S^z _{i\mu} \rangle$ is identical for all ${\cal N}$ orbitals in the orbitally degenerate ferromagnetic state. For the saturated ferromagnet, the magnetization $m$ is equal to the particle density $n$ for each orbital.

Due to orbital degeneracy, there are only two independent cases of interest corresponding to $\mu=\nu$ and  $\mu\neq\nu$. Following Eq. (\ref{c3e15}), the two corresponding coupled equations at the RPA level are:
\begin{equation}
\chi^{-+}_{\alpha\alpha}=\chi_0+\chi_{0}U\chi^{-+}_{\alpha\alpha}+({\cal N}-1)\chi_0J\chi^{-+}_{\beta\alpha},
\label{c3e7}
\end{equation}
\begin{equation}
\chi^{-+}_{\beta\alpha}=\chi_0J\chi^{-+}_{\alpha\alpha}+\chi_0U\chi^{-+}_{\beta\alpha}+({\cal N}-2)\chi_0J\chi^{-+}_{\beta\alpha},
\label{c3e8}
\end{equation}
solving which, we obtain
\begin{equation}
\chi^{-+}_{\alpha\alpha}=\frac{1}{\cal N}\left(\frac{\chi_0}{1-U^{+}\chi_0}\right) + \left(\frac{{\cal N}-1}{\cal N}\right)\left(\frac{\chi_0}{1-U^{-}\chi_0}\right),
\label{c3e9}
\end{equation}
\begin{equation}
\chi^{-+}_{\beta\alpha}=\frac{1}{\cal N}\left(\frac{\chi_0}{1-U^{+}\chi_0}\right) - \frac{1}{\cal N} \left(\frac{\chi_0}{1-U^{-}\chi_0}\right),
\label{c3e10}
\end{equation}
where the two interaction terms above are $U^{+}=U+({\cal N}-1)J$ and $U^{-}=U-J$. The propagators involve linear combinations of in-phase and out-of-phase modes with respect to the orbitals, representing gapless (acoustic) and gapped (optical) branches, respectively. The in-phase mode with effective interaction $U^+$ corresponds to the usual Goldstone mode (acoustic branch), while the out-of-phase mode with effective interaction $U^-$ yields gapped excitations (optical branch).\cite{mythesis}

\subsection{Quantum corrections and effective quantum parameter}
\begin{figure}
\begin{center}
\includegraphics[scale=0.8]{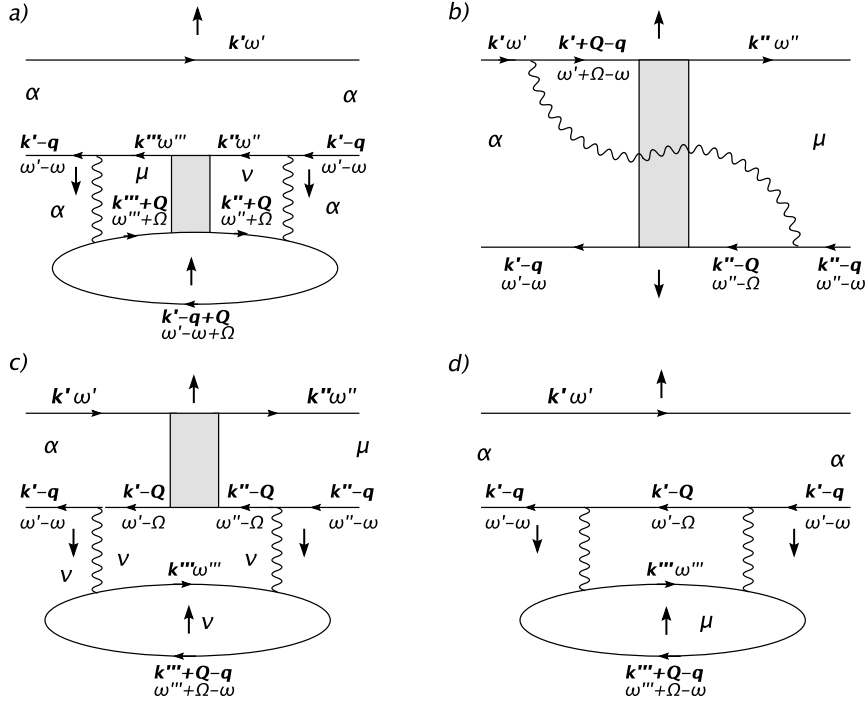}
\end{center}
\caption{First-order quantum corrections to the irreducible particle-hole propagator $\phi({\bf q},\omega)$ for the realistic ${\cal N}$-orbital model with arbitrary intra-orbital and inter-orbital Coulomb interactions $U$ and $J$.}
\label{c3f5}
\end{figure}

Diagrammatic contributions to the first order quantum correction $\phi^{(1)}$ for the orbitally asymmetric Hamiltonian (1) are shown in Fig. \ref{c3f5}. Structurally, they are similar to the O$(1/{\cal N})$ diagrams in the orbitally symmetric case,\cite{AS_2006} but the different orbital components with appropriate interaction terms are now considered separately. Diagrams (a) and (d) represent corrections to the irreducible propagator due to self-energy corrections, whereas diagrams (b) and (c) represent vertex corrections. The corresponding expressions are obtained as:\cite{mythesis}

\begin{eqnarray}
\phi^{(a)}({\bf q},\omega) &=& \sum_{\bf Q} \int \frac{d\Omega}{2\pi i} \left[(U^2+({\cal N}-1)J^2)\chi^{-+}_{\alpha\alpha}({\bf Q},\Omega)+2({\cal N}-1)UJ\chi^{-+}_{\beta\alpha}({\bf Q},\Omega)\right. \nonumber \\
&+& \left.({\cal N}-1)({\cal N}-2)J^2\chi^{-+}_{\beta\alpha}({\bf Q},\Omega)\right]\times\sum_{\bf k'} \left( \frac{1}{\epsilon_{\bf k'-q}^{\downarrow+}-\epsilon_{\bf k'}^{\uparrow-}+\omega-i\eta} \right)^2 \nonumber \\
&\times& \left( \frac{1}{\epsilon^{\uparrow+}_{\bf k'-q+Q} - \epsilon^{\uparrow-}_{\bf k'} + \omega - \Omega - i\eta} \right),
\label{a}
\end{eqnarray}
\begin{eqnarray}
\phi^{(b)}({\bf q},\omega)&=&-2\sum_{\bf Q}\int\frac{d\Omega}{2\pi i}\{U\Gamma^{-+}_{\alpha\alpha}({\bf Q},\Omega)+({\cal N}-1)J\Gamma^{-+}_{\alpha\beta}({\bf Q},\Omega)\} \nonumber \\
&\times& \sum_{\bf k'} \left( \frac{1}{\epsilon_{\bf k'-q}^{\downarrow+}-\epsilon_{\bf k'}^{\uparrow-}+\omega-i\eta} \right)\cdot\left( \frac{1}{\epsilon_{\bf k'-q+Q}^{\uparrow+}-\epsilon_{\bf k'}^{\uparrow-}+\omega-\Omega-i\eta} \right) \nonumber \\
&\times&\sum_{\bf k''}\left( \frac{1}{\epsilon_{\bf {k''-Q}}^{\downarrow+}-\epsilon_{\bf k''}^{\uparrow-}+\Omega-i\eta} \right)\cdot\left( \frac{1}{\epsilon_{\bf k''-q}^{\downarrow+}-\epsilon_{\bf k''}^{\uparrow-}+\omega-i\eta} \right),
\label{b}
\end{eqnarray}
\begin{eqnarray}
\phi^{(c)}({\bf q},\omega)&=&\sum_{\bf Q}\int\frac{d\Omega}{2\pi i}\{[U^2+({\cal N}-1)J^2]\Gamma^{-+}_{\alpha\alpha}({\bf Q},\Omega)+[2({\cal N}-1)JU \nonumber \\
&+&({\cal N}-1)({\cal N}-2)J^2]\Gamma^{-+}_{\alpha\beta}({\bf Q},\Omega)\} \nonumber \\
&\times& \left[ \sum_{\bf k'}\left( \frac{1}{\epsilon_{\bf k'-q}^{\downarrow+}-\epsilon_{\bf k'}^{\uparrow-}+\omega-i\eta}\right)\cdot\left(\frac{1}{\epsilon_{\bf k'-Q}^{\downarrow+}-\epsilon_{\bf k'}^{\uparrow-}+\Omega-i\eta} \right) \right]^2 \nonumber \\
&\times& \sum_{\bf k''}\left( \frac{1}{\epsilon_{\bf k''-q+Q}^{\uparrow+}-\epsilon_{\bf k''}^{\uparrow-}+\omega-\Omega-i\eta}\right),
\label{c}
\end{eqnarray}
\begin{eqnarray}
\phi^{(d)}({\bf q},\omega)&=&\sum_{\bf Q}\int\frac{d\Omega}{2\pi i}[U^2+({\cal N}-1)J^2]\sum_{\bf k'}\left[ \left( \frac{1}{\epsilon_{\bf k'-q}^{\downarrow+}-\epsilon_{\bf k'}^{\uparrow-}+\omega-i\eta}\right)^2 \right. \nonumber \\
&\times& \left. \left( \frac{1}{\epsilon_{\bf k'-Q}^{\downarrow+}-\epsilon_{\bf k'}^{\uparrow-}+\Omega-i\eta}\right)\right]\sum_{\bf k''}\left( \frac{1}{\epsilon_{\bf k''-q+Q}^{\uparrow+}-\epsilon_{\bf k''}^{\uparrow-}+\omega-\Omega-i\eta}\right), \nonumber \\
& & \label{d}
\end{eqnarray}
where the kernels $\Gamma^{-+}_{\alpha\alpha}$ and $\Gamma^{-+}_{\alpha\beta}$ for the spin-wave propagators are defined in terms of $\chi^{-+}_{\alpha\alpha}$ and $\chi^{-+}_{\alpha\beta}$, and from Eqs. (\ref{c3e9},\ref{c3e10}) are obtained as:
\begin{equation}
[\Gamma^{-+}_{\alpha\alpha}]_{\rm RPA}\equiv\frac{[\chi^{-+}_{\alpha\alpha}]_{\rm RPA}-\chi_0}{\chi_0^2}=\frac{1}{\cal N}\left[ \frac{U^{+}}{1-U^{+}\chi_0} + \frac{({\cal N}-1)U^{-}}{1-U^{-}\chi_0} \right],
\label{c3e24}
\end{equation}
\begin{equation}
[\Gamma^{-+}_{\alpha\beta}]_{\rm RPA}\equiv\frac{[\chi^{-+}_{\alpha\beta}]_{\rm RPA}}{\chi_0^2}=\frac{1}{\cal N}\left[ \frac{U^{+}}{1-U^{+}\chi_0} - \frac{U^{-}}{1-U^{-}\chi_0} \right].
\label{c3e25}
\end{equation}

To demonstrate the exact cancellation and hence the Goldstone mode for $q=0$, we note that the boson term (quantity in braces) in Eq. (\ref{b}) for $\phi^{(b)}$ can be expressed as:
\begin{eqnarray}
{ U\Gamma^{-+}_{\alpha\alpha}+({\cal N}-1)J\Gamma^{-+}_{\alpha\beta} } &=& 
\{[U^2+({\cal N}-1)J^2]\chi^{-+}_{\alpha\alpha} \nonumber \\ 
&+& [2({\cal N}-1)UJ+({\cal N}-1)({\cal N}-2)J^2]\chi^{-+}_{\alpha\beta} \}/\chi_0 \; ,
\label{bboson}
\end{eqnarray}
which is identical to the boson term in Eq. (\ref{a}) for $\phi^{(a)}$. The above identity is shown in the Appendix. Similarly, using Eqs. (\ref{c3e24}) and (\ref{c3e25}), the kernels in the boson term of Eq. (\ref{c}) for $\phi^{(c)}$ can be written in terms of $\chi^{-+}$ and $\chi_0$. Using the above substitutions in Eqs. (\ref{a})-(\ref{d}), and with $\epsilon^{\downarrow+}_{\bf k-q}-\epsilon^{\uparrow-}_{\bf k}=2\Delta$ for $q=0$, we obtain:
\begin{eqnarray}
& & \phi^{(1)}(q=0,\omega) = \phi^{(a)}+\phi^{(b)}+\phi^{(c)}+\phi^{(d)} \nonumber \\
&=& \sum_{\bf Q}\int\frac{d\Omega}{2\pi i}\left( \frac{1}{2\Delta+\omega-i\eta} \right)^2\sum_{\bf k'}\left( \frac{1}{\epsilon^{\uparrow+}_{\bf k'+Q}-\epsilon^{\uparrow-}_{\bf k'}+\omega-\Omega-i\eta} \right) \nonumber \\
&\times&[\{(U^2+({\cal N}-1)J^2)\chi^{-+}_{\alpha\alpha}+(2({\cal N}-1)UJ+({\cal N}-1)({\cal N}-2)J^2)\chi^{-+}_{\beta\alpha}\} \nonumber \\
&-2& \{(U^2+({\cal N}-1)J^2)\chi^{-+}_{\alpha\alpha}+(2({\cal N}-1)UJ+({\cal N}-1)({\cal N}-2)J^2)\chi^{-+}_{\beta\alpha}\} \nonumber \\
&+& \{(U^2+({\cal N}-1)J^2)(\chi^{-+}_{\alpha\alpha}-\chi_0)+(2({\cal N}-1)JU+({\cal N}-1)({\cal N}-2)J^2)\chi^{-+}_{\beta\alpha}\} \nonumber \\
&+&\{(U^2+({\cal N}-1)J^2)\chi_0 \}],
\label{c3e31}
\end{eqnarray}
which yields identically vanishing contribution for each spin-fluctuation mode ${\bf Q}$. We note that this mode-by-mode exact cancellation is quite independent of the spectral distribution of the spin-fluctuation spectrum between collective spin-wave and particle-hole Stoner excitations. Furthermore, the cancellation holds for all $\omega$, indicating no spin-wave amplitude renormalization, as expected for the saturated ferromagnet in which there are no quantum corrections to magnetization. 

We shall now obtain an effective quantum parameter which approximately determines the strength of the quantum corrections obtained above. Due to the uncorrelated nature of the inter-orbital spin fluctuations ($\langle S_{i\alpha} ^+ S_{i\beta} ^- \rangle = 0$), the contribution of the inter-orbital propagator $\chi^{-+}_{\alpha\beta}$ in Eq. (\ref{a}) is much smaller than the contribution of the intra-orbital propagator $\chi^{-+}_{\alpha\alpha}$, and hence only the contribution from the orbitally diagonal term $[U^2+({\cal N}-1)J^2]\chi^{-+}_{\alpha\alpha}$ essentially survives, leaving an overall factor $U^2+({\cal N}-1)J^2$ on carrying out the ${\bf Q},\Omega$ integration. Comparing with the corresponding factor $(U+({\cal N}-1)J)^2$ obtained for the equivalent single-orbital case (with identical exchange splitting) yields an overall relative factor of $[U^2+({\cal N}-1)J^2]/[U+({\cal N}-1)J]^2$. Similarly, the quantum corrections $\phi^{(b)}$, $\phi^{(c)}$ and $\phi^{(d)}$ also yield the same overall factor. Thus the total first-order quantum correction approximately bears an overall relative factor of $[U^2+({\cal N}-1)J^2]/[U+({\cal N}-1)J]^2$ compared to the equivalent single-orbital case. This relative factor thus plays the role of an effective quantum parameter which determines the strength of the correlation-induced quantum corrections in a multi-band metallic ferromagnet. This quantum parameter is exact in the orbitally independent limit $J/U\rightarrow 0$ where it approaches 1, and in the orbitally symmetric limit $J/U\rightarrow 1$ where it approaches $1/{\cal N}$. Also, the quantum parameter falls rapidly with Hund's coupling $J$, especially for large ${\cal N}$, highlighting essentially the role of orbital degeneracy and Hund's coupling in stabilizing metallic ferromagnetism by suppressing the quantum corrections. 

\subsection{Spin stiffness}
As the Goldstone mode is explicitly preserved in our approach, it allows investigation of correlation effects on spin stiffness and hence on the ferromagnetic stability with respect to long wavelength fluctuations. Here, we will evaluate the first-order quantum correction to spin stiffness exactly and then compare its $J$ dependence with that of the effective quantum parameter obtained above. This quantitative comparison will clearly show the usefulness of the effective quantum parameter. 

First-order quantum corrections to spin stiffness are derived by expanding $\phi^{(1)}({\bf q})$ for small ${\bf q}$ as in the two-orbital case.\cite{bhaskar} There is no quantum correction to the delocalization contribution $\langle {\mbox{\boldmath $\nabla$}}^2 \epsilon_{\bf k} \rangle $ in the spin-stiffness constant; only the exchange contribution in the spin stiffness is renormalized by the surviving second-order terms in 
$\delta \equiv \epsilon_{\bf k}-\epsilon_{{\bf k}-{\bf q}}$, and we obtain for the first-order quantum correction to spin stiffness:
\begin{eqnarray}
&D^{(1)}& = 2\Delta (U^{+}) \phi^{(1)} / q^2 \nonumber \\
&=& \frac{1}{d} \frac{U^+}{(2\Delta)^3} \sum_{\bf Q} \int \frac{d\Omega}{2\pi i} 
\left [ U_{\rm eff}^a({\bf Q},\Omega) \left ( \sum_{\bf k'} \frac{( {\mbox{\boldmath $\nabla$}} \epsilon_{\bf k'} )^2 }
{\epsilon_{\bf k' + Q}^{\uparrow +} - \epsilon_{\bf k'}^{\uparrow -} 
- \Omega - i\eta} \right ) \right . \nonumber \\
&-&  
\frac{2U_{\rm eff}^a({\bf Q},\Omega)} {\chi_0({\bf Q},\Omega)} \left ( \sum_{\bf k'} 
\frac{{\mbox{\boldmath $\nabla$}} \epsilon_{\bf k'}}
{\epsilon_{\bf k' + Q}^{\uparrow +} - \epsilon_{\bf k'}^{\uparrow -} 
- \Omega- i\eta} \right ) . \left ( \sum_{\bf k''} 
\frac{{\mbox{\boldmath $\nabla$}} \epsilon_{\bf k''}}
{\epsilon_{\bf k'' - Q}^{\downarrow +} - \epsilon_{\bf k''}^{\uparrow -} + \Omega- i\eta} \right )
\nonumber \\
&+& \frac{U_{\rm eff}^c ({\bf Q},\Omega)} {\chi_0 ^2 ({\bf Q},\Omega)} \left ( \sum_{\bf k'} 
\frac{1}{\epsilon_{\bf k' + Q}^{\uparrow +} - \epsilon_{\bf k'}^{\uparrow -} - \Omega - i\eta} \right ) 
\left ( \sum_{\bf k''} \frac{{\mbox{\boldmath $\nabla$}} \epsilon_{\bf k''}}
{\epsilon_{\bf k'' - Q}^{\downarrow +} - \epsilon_{\bf k''}^{\uparrow -} + \Omega- i\eta } \right )^2 
\nonumber \\
&+& 
\left . [U^2+({\cal N}-1)J^2] \left ( \sum_{\bf k'} \frac {1}{\epsilon_{\bf k' + Q}^{\uparrow +} - \epsilon_{\bf k'}^{\uparrow -} - \Omega- i\eta} \right ) \left ( \sum_{\bf k''} \frac{({\mbox{\boldmath $\nabla$}} \epsilon_{\bf k''})^2}
{\epsilon_{\bf k'' - Q}^{\downarrow +} - \epsilon_{\bf k''}^{\uparrow -} + \Omega- i\eta} \right ) \right ] , \nonumber \\
& & 
\label{c3e34}
\end{eqnarray}
where the effective interactions:
\begin{eqnarray}
U_{\rm eff}^a &=& [U^2+({\cal N}-1)J^2]\chi^{-+}_{\alpha\alpha}+[2({\cal N}-1)UJ+({\cal N}-1)({\cal N}-2)J^2]\chi^{-+}_{\beta\alpha}, \nonumber \\
U_{\rm eff}^c &=& U_{\rm eff}^a - (U^2 + ({\cal N}-1)J ^2) \chi_0 \; .
\end{eqnarray}

\begin{figure}
\begin{center}
\includegraphics[scale=0.8]{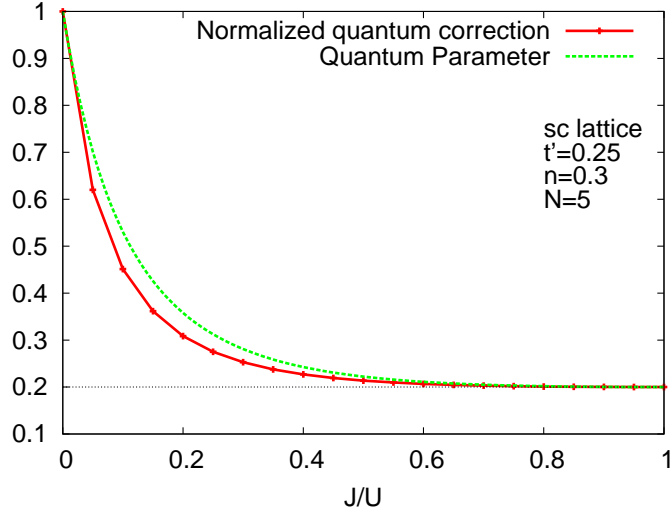}
\end{center}
\caption{The normalized first order quantum correction $D^{(1)}(J)/D^{(1)}(0)$ to spin stiffness calculated from Eq. (\ref{c3e34}) compared to the effective quantum parameter $[1+({\cal N}-1)(J/U)^2]/[1+({\cal N}-1)(J/U)]^2$ as a function of $J/U$ for ${\cal N}=5$ orbitals with fixed $U+({\cal N}-1)J=1.5W=18t$.}
\label{c3f8}
\end{figure}

The calculated quantum correction to spin stiffness, normalized so that it equals 1 for $J/U=0$, is plotted in Fig. \ref{c3f8} as a function of $J/U$ for ${\cal N}=5$ orbitals. Here we have considered the sc lattice, band filling $n=0.3$, $t'=0.25$, and fixed $U+({\cal N}-1)J=1.5W$. Also shown for comparison is the quantum parameter $'\hbar'$ obtained above, which is seen to be exact in the orbitally independent ($J/U\rightarrow 0$) and the orbitally symmetric ($J/U\rightarrow1$) limits, and it remains close to the calculated corrections even in the intermediate region, as expected from the uncorrelated nature of inter-orbital spin fluctuations.\cite{mythesis}

Including the quantum correction, the renormalized spin stiffness is then obtained as:
\begin{equation}
D = D^{(0)} - D^{(1)} \; ,
\end{equation}
where the bare-level (RPA) spin stiffness:
\begin{equation}
D^{(0)} = \frac{1}{d} \left [\frac{1}{2}
\langle {\mbox{\boldmath $\nabla$}}^2 \epsilon_{\bf k} \rangle  -
\frac{\langle ({\mbox{\boldmath $\nabla$}} \epsilon_{\bf k})^2 \rangle }{2\Delta} \right ]
\label{c3e33}
\end{equation}
involves two characteristic terms representing delocalization energy loss and exchange energy gain upon spin twisting.

\section{Application to iron}
In the previous section we obtained an effective quantum parameter $\frac{U^2+({\cal N}-1)J^2}{(U+({\cal N}-1)J)^2}$ in terms of the physically important parameters ${\cal N},U,J$ for a multi-orbital band ferromagnet. This quantum parameter determines, in analogy with $1/{\cal N}$ for the generalized Hubbard model and $1/S$ for quantum spin systems, the strength of quantum corrections to magnetic excitation energies. The quantum parameter is strongly suppressed by Hund's coupling $J$, and rapidly approaches the limiting value of $1/{\cal N}$, especially for large ${\cal N}$. 

This suggests that quantum corrections in a realistic multi-band ferromagnet with arbitrary ${\cal N},U,J$ can be conveniently investigated in terms of the orbitally symmetric Hubbard model:
\begin{equation}
H=-t\sum_{\langle ij\rangle,\sigma,\alpha} a^{\dag}_{i\sigma\alpha}a_{j\sigma\alpha} +\frac{U}{N}\sum_{i,\alpha,\beta}(a^{\dag}_{i\uparrow\alpha}a_{i\uparrow\alpha}a^{\dag}_{i\downarrow\beta}a_{i\downarrow\beta} + a^{\dag}_{i\uparrow\alpha}a_{i\uparrow\beta}a^{\dag}_{i\downarrow\beta}a_{i\downarrow\alpha})
\label{c4e1}
\end{equation}
and the inverse-degeneracy expansion scheme,\cite{AS_2006} with an effective number of orbitals $N$ such that $\frac{1}{N}=\frac{U^2+({\cal N}-1)J^2}{(U+({\cal N}-1)J)^2}$. In this section, we will apply this approach specifically to the case ${\cal N}=5$ with realistic parameters corresponding to ferromagnetic Fe. 
Accordingly, we consider a bcc lattice with dispersion:
\begin{equation}
\epsilon_{\bf k}=8t{\cos}({k_x a}){\cos}({k_y a}) {\cos}({k_z a}) + 2t' \sum_{\mu} {\cos}({2k_{\mu} a}),
\label{bcc_disp}
\end{equation}
where $t$ and $t'$ are the nearest and next-nearest neighbor hoppings, which yields bandwidth $W=16t$ (for $t'/t \le 0.5$). We have taken the lattice parameter $2a=2.87$\AA, $t'/t=0.5$, $t=0.2$ eV, $J/U\simeq 1/4$, and the interaction strength $U$ of the order of bandwidth $W$ as appropriate for a strongly correlated system. This yields $U\simeq W=16t=3.2$ eV and $J=0.8$ eV, which are close to the parameter values considered in band structure\cite{naito_JPSJ_2007} and constrained LDA\cite{lichtenstein_PRL_2001} studies for iron. With these parameters, and ${\cal N}=5$ corresponding to the five $3d$ orbitals in iron, the quantum parameter $'\hbar'=\frac{U^2+({\cal N}-1)J^2}{(U+({\cal N}-1)J)^2}\simeq\frac{1}{3}$, which corresponds to $N \simeq 3$ within the $N$-orbital Hubbard model. The choice of $t'/t=0.5$ is motivated by the fact that in a bcc lattice, the second neighbors are only about 15\% more distant than the nearest neighbors, and the hopping term $t_{ij}$ depends on the inter-site distance as $t_{ij}\propto 1/r_{ij}^5$,\cite{harrison_2004} which yields $t'/t\approx 0.5$.

\begin{figure}
\begin{center}
\includegraphics[scale=0.75]{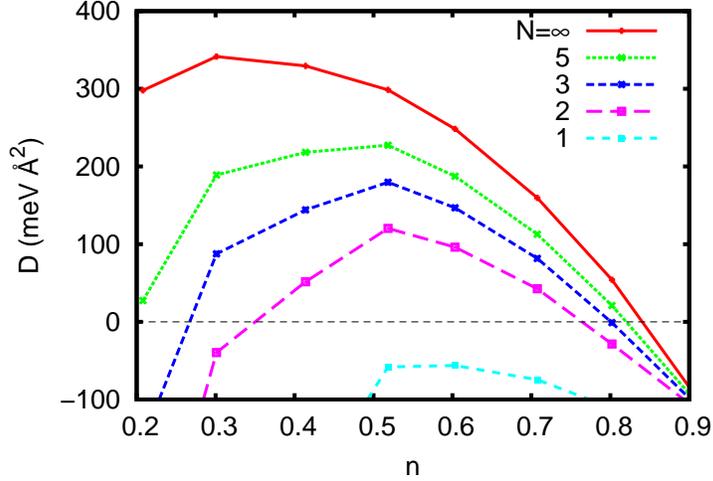}
\end{center}
\caption{The renormalized spin stiffness as a function of band filling for different effective number of orbitals $N$, evaluated for the bcc lattice with realistic bandwidth $W=16t=3.2$ eV, Coulomb interaction energy $U=W=3.2$ eV, and lattice parameter $2a=2.87$\AA\ for Fe. The measured value for Fe is 280 meV\AA$^2$.}
\label{c4f2}
\end{figure}

\subsection{Spin stiffness}
Fig. \ref{c4f2} shows the renormalized spin stiffness for the bcc lattice for different number of orbitals $N$. The spin stiffness is negative for $N=1$ and rapidly becomes positive with increasing $N$, indicating the strong role of orbital degeneracy in stabilizing ferromagnetism. The optimum filling occurs near $n=0.5$ for finite $N$. With realistic parameters taken for Fe as given above, the calculated values of the renormalized stiffness are close to the measured value of 280 
meV\AA$^2$ for iron obtained from neutron scattering studies.\cite{collins_PR_1969} This indicates that a simple multi-band Hubbard model with an effective number of orbitals to incorporate the quantum corrections provides a quantitative description of the spin stiffness for strongly correlated metallic ferromagnets such as Fe.

\subsection{Renormalized magnon dispersion}
The renormalized magnon energy $\omega_{\bf q}$ for mode ${\bf q}$ was obtained from the pole condition $[1-U{\rm Re}\phi({\bf q},-\omega_{\bf q})=0]$ in the total spin fluctuation propagator:
\begin{equation}
\chi^{-+}({\bf q},\omega)=\frac{\phi({\bf q},\omega)}{1-U\phi({\bf q},\omega)}
\label{ch4e6}
\end{equation}
where the irreducible particle-hole propagator:
\begin{equation}
\phi=\phi^{(0)} + \frac{1}{N}\phi^{(1)}
\end{equation}
up to first order in $1/N$. The numerical evaluation of the quantum correction $\phi^{(1)}$ by integrating over the intermediate $({\bf Q},\Omega)$ states has been discussed earlier.\cite{pandey_PRB_2007} Both collective and Stoner excitations are included. While the bare particle-hole propagator $\phi^{(0)}({\bf q},\omega)$ remains real in the relevant $\omega$ range, the quantum correction $\phi^{(1)}({\bf q},\omega)$ is complex for any finite $\omega$ due to the coupling with charge fluctuations, resulting in finite zero temperature magnon damping.\cite{sp_2008}

\begin{figure}
\begin{center}
\includegraphics[scale=0.75]{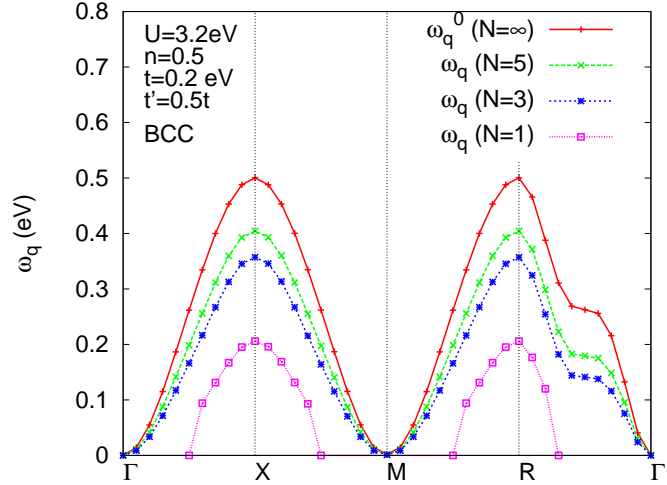}
\end{center}
\caption{Renormalized magnon energies for the bcc lattice for different number of orbitals $N$, showing a rapid crossover from negative-energy to positive-energy long wavelength modes as $N$ is increased from one, showing the strong role of orbital degeneracy in stabilizing metallic ferromagnetism.}
\label{c4f3}
\end{figure}

Fig. \ref{c4f3} shows the renormalized magnon energy dispersion for the bcc lattice for different $N$. The magnon energy exhibits a Goldstone mode at both $\Gamma$ and M, as they are equivalent points in our extended Brillouin zone which extends from $-\pi/a$ to $\pi/a$ in each direction, whereas the bcc lattice parameter is $2a$. We find that for $N=3$ and $5$, the magnon energy renormalization is nearly momentum independent in the $\Gamma$-X, X-M, and M-R directions. However, near $(\pi/2,\pi/2,\pi/2)$ between $\Gamma$-R, the magnon energy is softened relatively more strongly.

\subsection{Density of states and quasiparticle dispersion}
We now evaluate the renormalized electronic density of states (DOS) by incorporating the self energy correction due to electron-magnon coupling. A $\downarrow$-spin particle (energy $\epsilon_{\bf k}^{\downarrow} > \epsilon_F $) can decay into a magnon and a $\uparrow$-spin particle, resulting in considerable $\downarrow$-spin spectral weight transfer to just above the Fermi energy. Since we are considering a saturated ferromagnet with no $\downarrow$-spin density, an $\uparrow$-spin particle cannot decay into a $\downarrow$-spin particle due to spin conservation, and hence there are no quantum corrections at $T=0$ to the $\uparrow$-spin DOS.

The $\downarrow$-spin self-energy $\Sigma_{\downarrow}$ is calculated within an approximate resummation procedure which incorporates particle-particle correlations:\cite{hertz-edwards2}
\begin{equation}
\Sigma_{\downarrow}({\bf k},\omega) = \frac{\Sigma_{\downarrow}^{(0)}}{1-\Sigma_{\downarrow}^{(1)}({\bf k},\omega)/\Sigma_{\downarrow}^{(0)}},
\end{equation}
where $\Sigma_{\downarrow}^{(0)}=2\Delta=mU$ is the HF-level self energy, and the first order self energy:\begin{eqnarray}
\Sigma_{\downarrow}^{(1)}({\bf k},\omega) &=& \frac{1}{N}U^2\sum_{\bf Q}\int\frac{d\Omega}{2\pi i} \chi^{+-}_{\rm RPA}({\bf Q},\Omega)\left( \frac{1}{\omega-\Omega-\epsilon^{\uparrow+}_{{\bf k}-{\bf Q}}+i\eta} \right) \nonumber \\
&=& \frac{1}{N} mU^2\sum_{\bf Q}\frac{1}{\omega-\Omega^{0}_{\bf Q}-\epsilon_{{\bf k}-{\bf Q}}^{\uparrow+}+i\eta}.
\label{c4e9}
\end{eqnarray}

\begin{figure}
\begin{center}
\includegraphics[scale=0.75,angle=0]{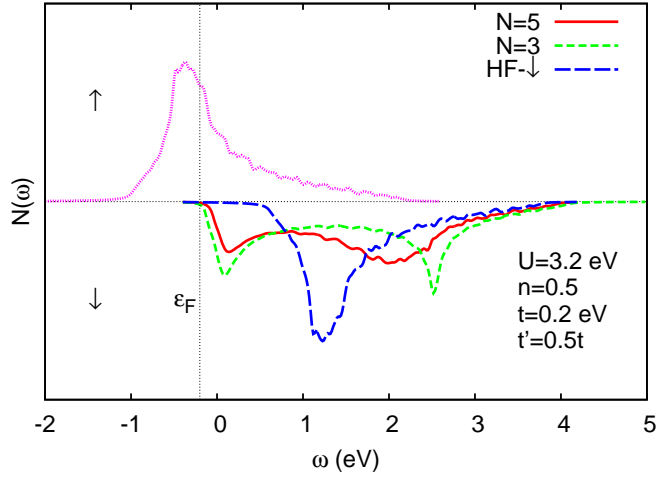}
\end{center}
\caption{The spin-resolved density of states for the bcc lattice at band filling $n=0.5$, showing the transfer of $\downarrow$-spin spectral weight due to correlation effects and the emergence of non-quasiparticle (NQP) states just above the Fermi energy.}
\label{c4f4}
\end{figure}

Fig. \ref{c4f4} shows the renormalized density of states for $N=$3 and 5 orbitals. The $\downarrow$-spin band HF DOS is also shown for comparison. The $\downarrow$-spin DOS is seen to be renormalized considerably with significant band narrowing, band shift, as well as spectral-weight transfer from the bare band to just above the Fermi energy.

\begin{figure}
\begin{center}
\vspace*{-60mm}
\hspace*{-20mm}
\includegraphics[scale=0.75,angle=0]{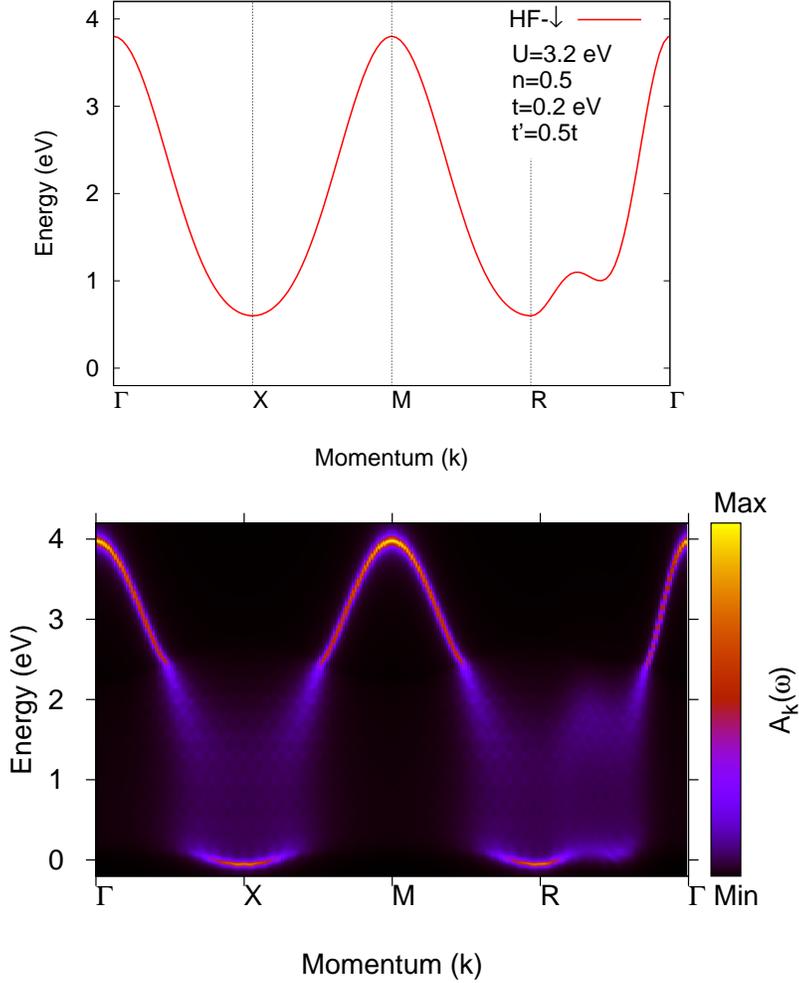}
\vspace{-20mm}
\end{center}
\caption{(color online) Intensity plot of the renormalized $\downarrow$-spin spectral function (bottom) along different symmetry directions in the Brillouin zone for the bcc lattice (with $N=5$). Also shown for comparison is the $\downarrow$-spin bare quasiparticle dispersion (top).}
\label{c4f5}
\end{figure}

Fig. \ref{c4f5} shows the energy momentum dispersion of $\downarrow$-spin electrons in terms of an intensity plot of the renormalized spectral function $A_{{\bf k}\downarrow}(\omega)$ obtained from the Green's function:
\begin{equation}
G_{\downarrow}({\bf k},\omega) = \frac{1}{G_0^{-1}({\bf k},\omega) - \Sigma_{\downarrow}({\bf k},\omega)}.
\end{equation}
The bare (HF) dispersion for the minority spin is also plotted for comparison. The renormalization is seen to be especially strong near the X and R points, indicating significant band flattening and mass renormalization, and also strong non-quasi-particle character of the low-energy minority-spin states near the Fermi energy, as further discussed below.

An important aspect in Fig. \ref{c4f4} is the emergence of new non-quasiparticle (NQP) states just above the Fermi energy at 0 K, corresponding to strongly incoherent spectral function near the X and R points in Fig. \ref{c4f5}. These NQP states are important in view of recent tunneling conductance measurements on Heusler alloy-based Magnetic Tunneling Junctions, where a strong suppression of spin polarization is observed with temperature.\cite{chioncel_PRL_2008} This suppression has been attributed to the emergence of NQP states at and below the Fermi energy due to correlation effects arising from the electron-magnon coupling. 

\subsection{Finite temperature spin dynamics}
With correlation effects incorporated, the irreducible particle-hole propagator $\phi({\bf q},\omega)$ yields renormalized magnon energies as obtained above, and also the effective spin couplings $J_{ij} = U^2 \phi_{ij}$ in an equivalent Heisenberg model after integrating out the fermionic degrees of freedom. At finite temperature, these renormalized magnon energies will determine magnetization reduction resulting from the electron-magnon coupling self energy which is explicitly of order $1/N$ as in Eq. (\ref{c4e9}). This is also similar to the low-temperature relative magnetization reduction of order $1/S$ in the spin-$S$ Heisenberg ferromagnet. In vew of this equivalence with an effective spin-$S$ Heisenberg model with $N=2S$, we will therefore calculate the finite temperature magnetization from the self-consistent Callen equation for the spin-$S$ Heisenberg model:\cite{callen_1963}
\begin{equation}
\langle S_z \rangle_T = \frac{(S-\Phi)(1+\Phi)^{2S+1}+(S+1+\Phi)\Phi^{2S+1}}{(1+\Phi)^{2S+1}-\Phi^{2S+1}},
\label{c4e18}
\end{equation}
where the magnon amplitude:

\begin{equation}
\Phi = \sum_{\bf Q}\frac{1}{e^{\beta {\tilde \Omega_{\bf Q}}}-1},
\label{c4e19}
\end{equation}
in terms of the thermally renormalized magnon energies:

\begin{equation}
{\tilde {\Omega}}_{\bf Q}  =  \Omega_{\bf Q} \langle S_z \rangle_T / \langle S_z \rangle_0. 
\label{c4e20}
\end{equation}
Solving the above three coupled equations self-consistently with $S=N/2$ yields the temperature dependence of the magnetization. While the finite-temperature electronic density and magnetization corrections, which renormalize the $\omega$ term in the magnon propagator, are incorporated in this calculation through the thermal renormalization of magnon energies, it neglects intrinsic finite-temperature renormalization effects on the spin couplings.

\begin{figure}
\begin{center}
\includegraphics[scale=0.6,angle=0]{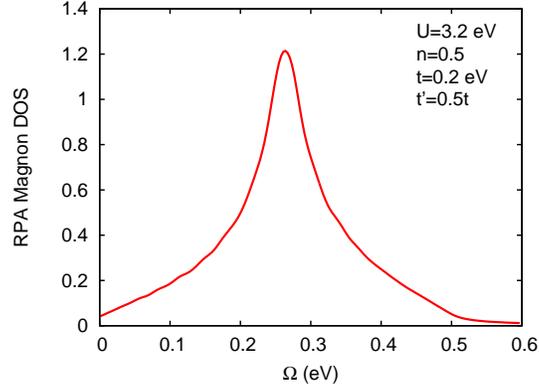}
\end{center}
\caption{The RPA-level magnon DOS for the bcc lattice, showing a peak at the dominant mode energy $\Omega^*_{\rm RPA}=0.26$ eV.}
\label{c4f8}
\end{figure}

\begin{figure}
\begin{center}
\includegraphics[scale=0.7,angle=0]{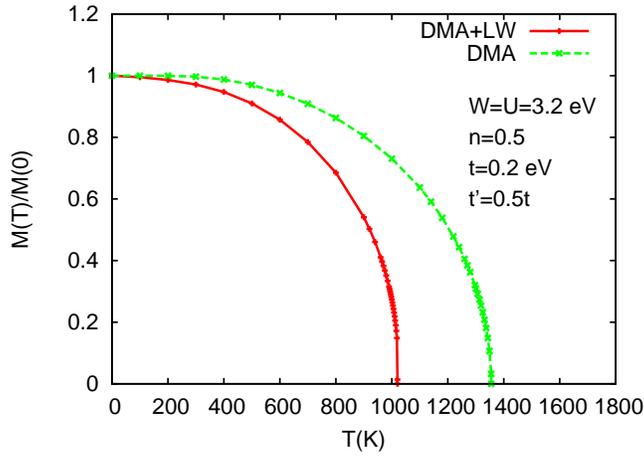}
\end{center}
\caption{Including long wavelength modes in the self-consistent Callen scheme for magnetization as a function of temperature results in a nearly 25\% suppression of $T_C$, as seen by comparison with the dominant mode approximation (DMA), highlighting the significant role of long wavelength spin fluctuations in determining the Curie temperature.}
\label{c4f10}
\end{figure}

Instead of performing the computationally intensive ${\bf Q}$ summation in Eq. (\ref{c4e19}) exactly, it is convenient to break it into two parts, corresponding to long- and short-wavelength contributions:
\begin{equation}
\Phi = \sum_{\bf Q}\frac{1}{e^{\beta {\tilde \Omega_{\bf Q}}}-1} = \sum_{Q<\Lambda}\frac{1}{e^{\beta {\tilde \Omega_{\bf Q}}}-1} + \sum_{Q>\Lambda}\frac{1}{e^{\beta {\tilde \Omega_{\bf Q}}}-1},
\label{c4e22}
\end{equation}
where $\Lambda$ represents a momentum space cut-off for the long wavelength modes. The small-$Q$ contribution can be evaluated explicitly by integration, whereas the large-$Q$ contribution can be approximated by a single term corresponding to the ``dominant mode energy", since the magnon DOS exhibits a pronounced peak at this energy corresponding to short wavelength modes. Thus, $\Phi=\Phi_{\rm LW}+\Phi_{\rm DMA}$, corresponding to the long wavelength contribution and the dominant mode approximation. The evaluation of $\Phi_{\rm LW}$ is discussed below.

For small-$Q$ modes, the magnon energy ${\tilde \Omega_{\bf Q}} = {\tilde D}Q^2$ in terms of the thermally renormalized spin stiffness:
\begin{equation}
{\tilde D}=D\langle S_z \rangle_T / \langle S_z \rangle_0
\end{equation}
at temperature $T$, where $D$ incorporates the zero-temperature quantum corrections discussed above. Using the transformation $x={\tilde D}Q^2/k_{\rm B}T$, we obtain:
\begin{equation}
\Phi_{\rm LW}= \frac{4}{(2\pi/a)^3} \int_0^\Lambda \frac{4\pi Q^2 dQ}{e^{\beta{\tilde D}Q^2} - 1} =\frac{4a^3}{(2\pi)^2}\left(\frac{k_{\rm B}T}{\tilde D}\right)^{3/2} \int_0^{x_c} dx \frac{\sqrt x}{e^x -1},
\label{c4e23}
\end{equation}
where $x_c = {\tilde D}\Lambda^2/k_{\rm B}T$ corresponds to the momentum cut-off $\Lambda$. For low temperatures ($k_{\rm B}T\ll {\tilde D}\Lambda^2$), the cutoff $x_c \gg 1$, and the $x$-integral is nearly temperature independent, which yields the well-known Bloch $T^{3/2}$ law for the decay in magnetization.

While including the long wavelength contribution, it should be noted that apart from $(0,0,0)$, there are twelve other points in the extended Brillouin zone which correspond to the Goldstone mode. These points are located at $(\pm\pi,\pm\pi,0), (0,\pm\pi,\pm\pi)$, and $(\pm\pi,0,\pm\pi)$, and each of these points is associated with a quadrant of a sphere in momentum space. Therefore, there are effectively four points in the Brillouin zone around which $\Omega_{\bf Q}=DQ^2$, and hence the factor 4 in Eq. (\ref{c4e23}).

Fig. \ref{c4f10} shows the finite temperature magnetization evaluated by solving the three coupled equations  (\ref{c4e18}-\ref{c4e20}) with $S=3/2$ and the renormalized magnon energies obtained in subsection B. The results are shown for both the dominant mode approximation (DMA) and after including the long wavelength (LW) contribution. The momentum cut-off for long wavelength modes was fixed at $\Lambda a =1$. The Curie temperature obtained within the DMA is approximately 1350 K. Including the long wavelength contribution reduces the Curie temperature to 1020 K. An important result of our analysis is that long wavelength modes yield a significant (nearly 25\%) reduction in the Curie temperature. This comparison provides a useful quantitative measure of the overestimation of the Curie temperature in calculations which neglect long wavelength modes such as in DMFT studies for iron.\cite{lichtenstein_PRL_2001}

\section{Application to ${\rm {\bf Co_2MnSi}}$}
As another application, we consider the Heusler alloy ${\rm Co_2MnSi}$, which, owing to its half-metallicity and high $T_C$ has attracted considerable attention in recent years due to possible applications in fabricating spintronics-based devices.\cite{buczek,galanakis,co2mnsiexp2,co2mnsiexp3} The key feature being exploited is the 100\% spin polarization arising due to the gap in the minority-spin DOS at the Fermi level. However, a rapid depolarization is seen with increasing temperature, which has been attributed to the emergence of non-quasiparticle states in the minority-spin band at the Fermi level.\cite{chioncel_PRL_2008} These non-quasiparticle states offer strong evidence of the vital role played by electron-magnon interactions in these systems. Indeed, as was shown in Section VI, the electron-magnon self-energy results in a transfer of spectral weight in the minority-spin band near the Fermi energy. In the following, we will investigate correlation effects on spin stiffness of ${\rm Co_2MnSi}$ within a simplified picture.

\begin{figure}
\begin{center}
\includegraphics[scale=0.5,angle=0]{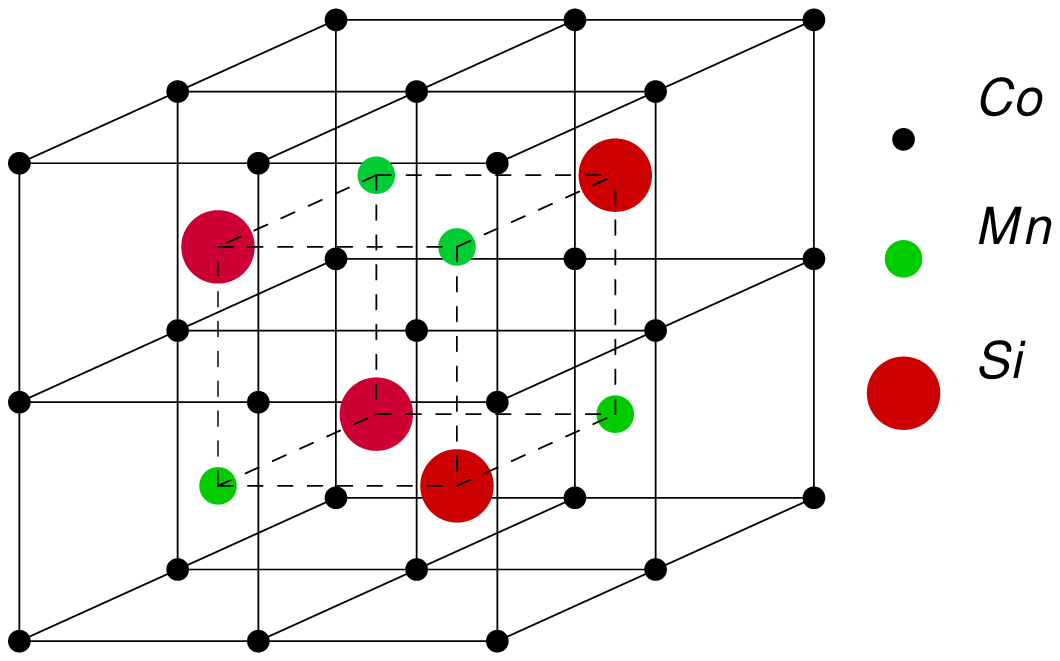}
\end{center}
\caption{Lattice structure of ${\rm Co_2MnSi}$.}
\label{co2mnsi_lattice}
\end{figure}

The lattice structure of ${\rm Co_2MnSi}$ is shown in Fig. \ref{co2mnsi_lattice}. The Co atoms occupy the corners of a simple cubic lattice, and the centers of the cubes are occupied by Si and Mn atoms alternately. Since the $sp$ electrons of Si do not participate in the electronic and magnetic properties of the system as they lie much below the Fermi level,\cite{galanakis} we consider only the Co and Mn atoms, the hybridization of whose orbitals results in the minority-spin band gap at $\epsilon_F$.\cite{galanakis} Thus the lattice effectively consists of an admixture of the sc and bcc lattices, and so we consider the following simplified dispersion in our analysis:

\begin{figure}
\begin{center}
\hspace*{-20mm}
\includegraphics[scale=0.7,angle=0]{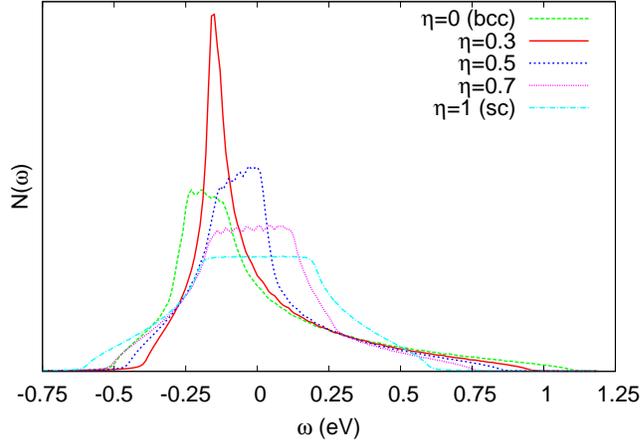}
\end{center}
\caption{The electronic DOS corresponding to the band dispersion given by Eq. (\ref{eqeta}) for different values of the tunable parameter $\eta$. Here $t=0.1$ eV and $t'/t=0.5$.}
\label{co2mnsidos}
\end{figure}

\begin{figure}
\begin{center}
\hspace*{-20mm}
\includegraphics[scale=0.35,angle=0]{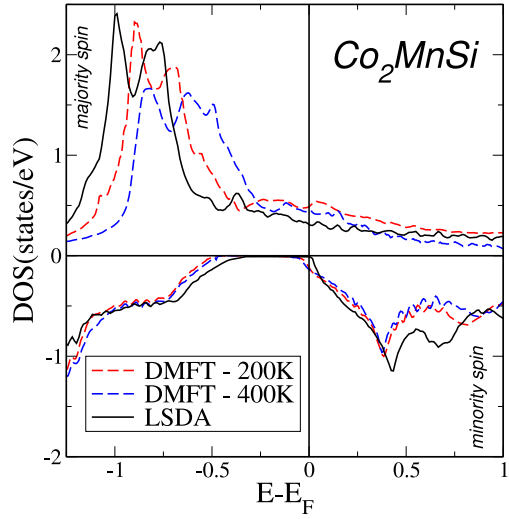}
\end{center}
\caption{Electronic DOS of ${\rm Co_2MnSi}$ as obtained from LDA+DMFT studies.\cite{chioncel_PRL_2008}.}
\label{co2mnsi_chioncel}
\end{figure}

\begin{figure}
\begin{center}
\hspace*{-20mm}
\includegraphics[scale=0.7,angle=0]{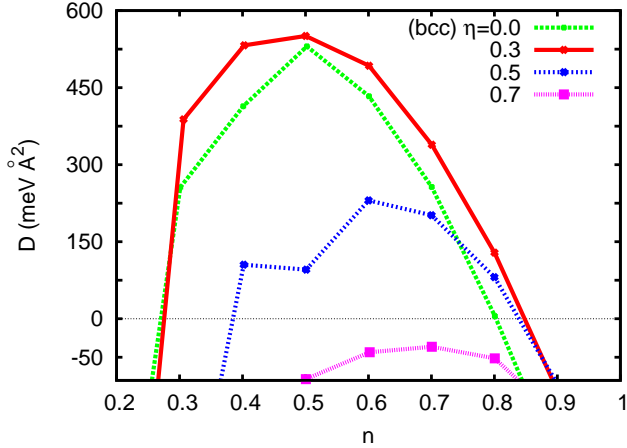}
\end{center}
\caption{Renormalized spin stiffness corresponding to the band dispersion given by Eq. (\ref{eqeta}) with $t=0.15$ eV and $t'/t=0.5$, and the effective quantum parameter $1/N=1/3$. The spin stiffness passes through a maximum at around $\eta=0.3$, with a crossover to negative values around $\eta \sim 0.7$.}
\label{co2mnsistiff}
\end{figure}

\begin{equation}
\epsilon_{\bf k} = (1-\eta)\epsilon_{\bf k}^{\rm bcc} + \eta\epsilon_{\bf k}^{\rm sc},
\label{eqeta}
\end{equation}
where $\eta$ is a tunable parameter which controls the relative strengths of the bcc and sc characters of the dispersion, and the bcc and sc band dispersions are given by:

\begin{equation}
\epsilon_{\bf k}^{\rm bcc}=8t{\cos}({k_x a}){\cos}({k_y a}) {\cos}({k_z a}) + 2t' \sum_{\mu} {\cos}({2k_{\mu} a}),
\label{bcc_disp2}
\end{equation}

\begin{equation}
\epsilon_{\bf k}^{\rm sc}= 2t({\cos}({2k_x a})+{\cos}({2k_y a}) + {\cos}(2{k_z a})).
\label{sc_disp}
\end{equation}

The lattice parameter for ${\rm Co_2MnSi}$ is 5.654 \AA,\cite{webster} nearly twice that of iron. For $t=0.1$ eV and $t'/t=0.5$, Fig. \ref{co2mnsidos} shows the electronic density of states for different $\eta$ values. The transition from a purely bcc band for $\eta=0$ to a purely sc band for $\eta=1$ is marked by a sharp peak near the lower band edge for $\eta \sim 0.3$, which should be particularly favourable for ferromagnetism. While the spectral distribution is qualitatively similar to the majority-spin DOS obtained from first principles calculations\cite{chioncel_PRL_2008} shown in Fig. \ref{co2mnsi_chioncel}, the bandwidth ($\sim$ 1.2 eV) is underestimated by roughly 50


We have again taken the interaction parameters $U/t=16$ and $J/U=1/4$ as in Section IV, which are similar to those considered in LSDA+DMFT studies.\cite{chioncel_PRL_2008} With the effective quantum parameter $'\hbar'=1/N=\frac{1}{3}$, the renormalized spin stiffness as a function of band filling $n$ is shown in Fig. \ref{co2mnsistiff} for different $\eta$ values. Here the hopping term is taken as $t=0.15$eV so that the majority-spin bandwidth of about 2eV matches with the first-principles calculations shown above.\cite{chioncel_PRL_2008} The spin stiffness maximum at $\eta \sim 0.3$ corresponds to the sharply peaked electronic DOS in Fig. \ref{co2mnsidos}, and the peak value of about 600 meV\AA$^2$ compares favourably with the stiffness value 575 meV\AA$^2$ obtained\cite{co2mnsiexp2,co2mnsiexp3} for the $\rm L2_1$ ordered phase (structure shown in Fig. \ref{co2mnsi_lattice}). For the ${\rm B_2}$ phase of $\rm Co_2 MnSi$, where the Mn and Si atoms are randomly disordered, a much lower spin stiffness of 324 meV\AA$^2$ was obtained.\cite{co2mnsiexp2} 

\section{Realistic multi-band systems}
The quantum corrections analysis discussed so far was restricted to an ${\cal N}$-fold degenerate Hubbard model. In the following, we present an extension of our Goldstone-mode preserving approach to realistic multi-band models with non-degenerate orbitals and inter-orbital hopping, as obtained from first principles calculations. An approximate scheme for the evaluation of the correlation-induced quantum corrections for a general multi-band model is discussed below. 

We consider the following multi-band Hamiltonian for realistic systems which includes non-degenerate orbitals and inter-orbital hopping:
\begin{equation}
H=-\sum_{\langle ij \rangle\mu\nu}t^{\mu\nu}_{ij}a^\dagger_{i\mu}a_{j\nu}-\sum_{i,\mu,\nu} U_{\mu\nu} {\bf S}_{i\mu}\cdot{\bf S}_{i\nu} 
=\sum_{{\bf k},\mu\nu} \epsilon_{\bf k} ^{\mu\nu} a_{{\bf k}\mu} ^\dagger a_{{\bf k}\nu} -\sum_{i,\mu,\nu} U_{\mu\nu} {\bf S}_{i\mu}\cdot{\bf S}_{i\nu} 
\label{hm}
\end{equation}
where the generalized interaction matrix $U_{\mu\nu}$ includes the intra-orbital Coulomb interaction $U_\mu$ for $\mu=\nu$ and the inter-orbital Hund's exchange $J_{\mu\nu}$ between orbitals $\mu$ and $\nu$ for $\mu\neq\nu$. 

As in the ${\cal N}$-fold orbitally degenerate case,\cite{bhaskar} the transverse spin-fluctuation propagator components are expressed in terms of the irreducible particle-hole propagator as:
\begin{equation}
[\chi^{-+}({\bf q},\omega)]_{\mu\nu} = [\phi({\bf q},\omega)]_{\mu\nu} + \sum_{\mu'\nu'}[\phi({\bf q},\omega)]_{\mu\mu'}[U]_{\mu'\nu'}[\chi^{-+}({\bf q},\omega)]_{\nu'\nu}
\label{chi_eqn}
\end{equation}
which can equivalently be written in a matrix form in the orbital basis:
\begin{equation}
[\chi^{-+}({\bf q},\omega)] = \frac{[\phi({\bf q},\omega)]}{{\bf 1} - [U][\phi({\bf q},\omega)]}.
\label{thirtyone}
\end{equation}

\begin{figure}
\begin{center}
\vspace*{-10mm}
\includegraphics[scale=0.5,angle=0]{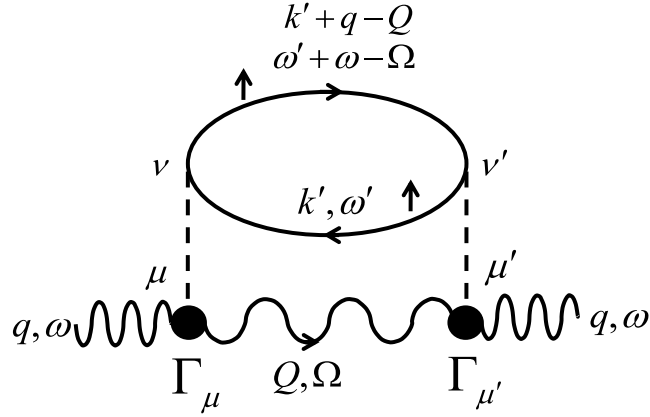}
\vspace{-30mm}
\end{center}
\caption{The spin-charge coupling structure of the first-order quantum correction $\phi^{(1)}({\bf q},\omega)$ in the general multi-band case with non-degenerate orbitals and inter-orbital hopping.}
\label{phi_multiband}
\end{figure}

Correlation-induced quantum corrections to the irreducible particle-hole propagator can be obtained by systematically incorporating self energy and vertex corrections as in the orbitally degenerate case. An approximate diagrammatic scheme for evaluating the first-order quantum correction $\phi^{(1)}({\bf q},\omega)$ in the general multi-band case is shown in Fig. \ref{phi_multiband}, which follows from the spin-charge coupling structure of the quantum correction and the magnon self energy for the saturated ferromagnet in the ${\cal N}$-orbital Hubbard model.\cite{sp_2008} The corresponding expression for the first-order quantum correction is obtained as:

\begin{equation}
[\phi^{(1)} ({\bf q},\omega)]_{\mu\mu'} = \sum_{\bf k,Q} \int \frac{d\Omega}{2\pi i} \; 
\Gamma_\mu \; 
[\chi^{-+} _{\mu\mu'}({\bf Q},\Omega) \; 
U^{\rm eff} _{\mu\mu'} ({\bf k};{\bf q-Q},\omega-\Omega)] \; 
\Gamma_{\mu'} 
\label{qc_multiband}
\end{equation}
where the minority-spin effective interaction $U^{\rm eff} _{\mu\mu'} = U_{\mu\nu} \Pi^0 _{\nu\nu'} U_{\nu'\mu'}$ involves the exchange of charge fluctuation propagator in the majority-spin bands:
\begin{eqnarray}
\Pi^0 _{\nu\nu'} ({\bf k};{\bf q-Q},\omega-\Omega) &=& 
i\int \frac{d\omega'}{2\pi} [G^0 _\uparrow ({\bf k+q-Q},\omega'+\omega-\Omega)]_{\nu\nu'}
[G^0 _\uparrow ({\bf k},\omega')]_{\nu'\nu} \nonumber \\
&=& \frac{\delta_{\nu\nu'}}
{\epsilon_{\bf k-q+Q}^{\nu\uparrow +} - \epsilon_{\bf k}^{\nu\uparrow -} + \omega - \Omega - i \eta } 
\end{eqnarray}
if orbital (band) mixing is neglected. Furthermore, in the orbitally degenerate case, the effective interaction term yields the factor $[U^2 + ({\cal N}-1)J^2]$, and the effective quantum parameter is recovered as earlier. This correlation-induced coupling between the spin and charge fluctuations represents scattering of a magnon into intermediate spin-excitation states accompanied by charge fluctuations in the majority spin band. These intermediate states include both the sharp magnon excitations and the Stoner excitations spread over the Stoner continuum. Finite imaginary part of this magnon self energy due to the gapless charge excitations results in finite magnon damping at zero temperature even for low-energy magnon modes lying within the Stoner gap.

The spin-charge interaction vertex above is given by: 
\begin{equation}
\Gamma_\mu ({\bf k};{\bf q},\omega;{\bf Q},\Omega) = \left (\chi^0 _\mu ({\bf k};{\bf q},\omega) - \frac{1}{2\Delta '({\bf q},\omega;{\bf Q},\Omega)} \right ) \; ,
\label{int_vertex}
\end{equation}
where (in the absence of band mixing)
\begin{equation}
\chi^0 _\mu({\bf k};{\bf q},\omega) \equiv \frac{1}
{\epsilon_{\bf k-q}^{\mu\downarrow +} - \epsilon_{\bf k}^{\mu\uparrow -} + \omega - i\eta} \; ,
\end{equation}
and
\begin{equation}
\frac{1}{2\Delta ' _\mu ({\bf q},\omega;{\bf Q},\Omega)} \equiv
\sum_{\bf k'} \chi^0 _\mu({\bf k'; q},\omega) \chi^0 _\mu({\bf k'; Q},\Omega) / \chi^0 _\mu({\bf Q},\Omega)\; ,
\end{equation}
which generally has weak momentum dependence due to the averaging over momentum ${\bf k'}$. For $q=0$, both terms in Eq. (\ref{int_vertex}) reduce to $1/(2\Delta + \omega)$. The spin-charge interaction vertex $\Gamma$ and the magnon self energy therefore vanish identically, and the Goldstone mode is thus explicitly preserved. For small $q$, $\Gamma^2 \sim ({\bf q}.{\mbox{\boldmath $\nabla$}} \epsilon_{\bf k})^2$, indicating short-range interaction. Also, the spin-charge coupling results in a quantum correction only to the exchange contribution to the spin stiffness as required; quantum corrections to the delocalization contribution of the type $({\bf q}.{\mbox{\boldmath $\nabla$}})^2 \epsilon_{\bf k}$ cancel exactly.\cite{AS_2006}

The RPA-level description used above in Eq. (\ref{qc_multiband}) is obtained by replacing the irreducible particle-hole propagator $[\phi]_{\mu\nu}$ in Eq. (\ref{chi_eqn}) by the bare particle-hole propagator:
\begin{equation}
[\chi_0({\bf q},\omega)]_{\mu\nu}=i\int_{-\infty}^{\infty}\frac{d\omega'}{2\pi}\sum_{\bf k'}[G^0_{\uparrow}({\bf k'},\omega')]_{\mu\nu}[G^0_{\downarrow}({\bf k'}-{\bf q},\omega'-\omega)]_{\mu\nu}
\end{equation}
in terms of the HF-level Green's function matrices in orbital space: 

\begin{equation}
[G^0_{\sigma}({\bf k},\omega)] = [\omega {\bf 1} - H_\sigma ^{\rm HF} ({\bf k})]^{-1} \; ,
\end{equation}
which are no longer diagonal due to orbital mixing by the hopping term. The HF-level Hamiltonian matrix in momentum-orbital-spin space:
\begin{equation}
[H^{\rm HF} _\sigma ({\bf k})] = [\epsilon_{\bf k}^{\mu\nu} ] - 2 {\bf \Delta}_\mu \cdot {\bf S}_\mu 
\end{equation}
in terms of the bare band energies and the self-consistent mean fields for orbital $\mu$:
\begin{equation}
{\bf \Delta}_{\mu}=U_\mu \langle {\bf S}_{\mu} \rangle + \sum_\nu J_{\mu\nu} \langle {\bf S}_{\nu} \rangle
\end{equation}
involving both the intra-orbital Hubbard interaction and the inter-orbital Hund's coupling. Assuming the magnetization to lie along the ${\hat z}$ direction, the exchange splitting is obtained as
$2\Delta_{\mu}=U_\mu m_\mu + \sum_{\nu} J_{\mu\nu}m_{\nu}$, where $m_{\mu}$ is the magnetization in orbital $\mu$.

\section{Conclusions}
The correlated motion of electrons in metallic ferromagnets was investigated in terms of a realistic Hubbard model with ${\cal N}$-fold orbital degeneracy and arbitrary intra-orbital Coulomb interaction $U$ and inter-orbital Hund's exchange $J$. A spin-rotationally symmetric and non-perturbative scheme was developed to study correlation-induced quantum corrections beyond the RPA, wherein self energy and vertex corrections were incorporated systematically so that the Goldstone mode is explicitly preserved order by order. An effective quantum parameter $'\hbar'=\frac{U^2+({\cal N}-1)J^2}{(U+({\cal N}-1)J)^2}$ was obtained which determines, in analogy with $1/S$ for quantum spin systems and $1/N$ for the $N$-orbital Hubbard model, the strength of quantum corrections to spin stiffness and magnon energies. The rapid suppression of this quantum parameter with Hund's coupling $J$, especially for large ${\cal N}$, provides fundamental insight into the phenomenon of strong stabilization of metallic ferromagnetism by orbital degeneracy and Hund's coupling.

The above approach was illustrated for the case of ferromagnetic iron. Electronic and magnetic properties were investigated on a bcc lattice with realistic parameters. The electronic spectral function renormalization was seen to be especially strong near the X and R points, indicating significant band flattening and mass renormalization, and also strong non-quasi-particle character of the low-energy minority-spin states near the Fermi energy. With the same set of parameters, both the calculated spin stiffness and Curie temperature values obtained were in quantitative agreement with measurements. An important finding of our explicitly Goldstone mode preserving scheme was the result that including the contribution of long wavelength modes yielded a nearly $\sim25\%$ reduction in the calculated Curie temperature, which should be of interest in view of the significant over-estimation of the Curie temperature in approaches where only local magnetic excitations are included. Another illustration was provided for the half metallic Heusler alloy $\rm Co_2 Mn Si$ and the calculated spin stiffness was again in agreement with measured values. 

Finally, an outline was presented for calculating the correlation-induced quantum corrections for a generic multi-band metallic ferromagnet including realistic band-structure features of non-degenerate orbitals and inter-orbital hopping as obtained from LDA calculations. This extension drew upon the spin-charge coupling structure, which explicitly preserves the Goldstone mode, highlights the role of majority-spin charge fluctuations on spin dynamics, and also explicitly yields finite zero-temperature magnon damping due to decay into longer wavelength modes accompanied with internal charge excitations. This provides an important advantage over other approaches which incorporate correlation effects in terms of an effective Heisenberg model only. 

\section{Appendix}

The identity used in Eq. (\ref{bboson}) is derived here. From Eqs. (\ref{c3e24},\ref{c3e25}) and with $U=[U^+ + ({\cal N}-1)U^-]/{\cal N}$ and $J=[U^+ - U^-]/{\cal N}$, we obtain:
\begin{equation}
U\Gamma^{-+}_{\alpha\alpha}+({\cal N}-1)J\Gamma^{-+}_{\alpha\beta} = \frac{1}{\cal N}\left[    \frac{(U^+)^2}{1-U^+\chi_0} + \frac{({\cal N}-1)(U^-)^2}{1-U^-\chi_0}    \right],
\end{equation}
which can be written in terms of a superposition:
\begin{eqnarray}
\frac{1}{\cal N} \left[ \frac{(U^+)^2}{1-U^+\chi_0} + \frac{({\cal N}-1)(U^-)^2}{1-U^-\chi_0} \right] &=& \frac{\cal A}{\cal N}\left[\frac{1}{1-U^+\chi_0} + \frac{{\cal N}-1}{1-U^-\chi_0}\right] \nonumber \\
&+& \frac{\cal B}{\cal N}\left[\frac{1}{1-U^+\chi_0} - \frac{1}{1-U^-\chi_0}\right]
\end{eqnarray}
of the two functions in Eqs. (\ref{c3e9}) and (\ref{c3e10}). Solving for ${\cal A}$ and ${\cal B}$ yields:
\begin{eqnarray}
{\cal A} &=& U^2+({\cal N}-1)J^2 \nonumber \\
{\cal B} &=& 2({\cal N}-1)UJ+({\cal N}-1)({\cal N}-2)J^2,
\end{eqnarray}
so that in terms of $\chi^{-+}_{\alpha\alpha}$ and $\chi^{-+}_{\alpha\beta}$ from Eqs. (\ref{c3e9}) and (\ref{c3e10}), we obtain:
\begin{eqnarray}
U\Gamma^{-+}_{\alpha\alpha}+({\cal N}-1)J\Gamma^{-+}_{\alpha\beta} &=& 
\{[U^2+({\cal N}-1)J^2]\chi^{-+}_{\alpha\alpha} \nonumber \\ 
&+& [2({\cal N}-1)UJ+({\cal N}-1)({\cal N}-2)J^2]\chi^{-+}_{\alpha\beta} \}/\chi_0.
\label{appendix}
\end{eqnarray}

\end{document}